\documentclass[preprintnumbers,showkeys,amsmath,amssymb]{revtex4}
\usepackage{graphicx}
\usepackage{dcolumn}
\usepackage{bm}
\begin{document}

\preprint{}

\title{Entanglement concentration and purification of two-mode squeezed microwave photons in circuit QED\footnote{Published in Annals of Physics \textbf{391}, 112-119 (2018).}}

\author{Hao Zhang$^{1}$, Ahmed Alsaedi$^{2}$, Tasawar  Hayat$^{2,3}$, Fu-Guo
Deng$^{\text{1,2}}$\footnote{Corresponding author:
fgdeng@bnu.edu.cn}}

\address{$^{1}$Department of Physics, Applied Optics Beijing Area Major Laboratory,
Beijing Normal University, Beijing 100875, China\\
$^{2}$NAAM Research Group, Department of Mathematics, King Abdulaziz
University, Jeddah 21589, Saudi Arabia\\
$^{3}$Department of Mathematics, Quaid-I-Azam University, Islamabad
44000, Pakistan }

\begin{abstract}
We present a theoretical proposal for a physical implementation of
entanglement concentration and purification protocols for two-mode
squeezed microwave photons in circuit quantum electrodynamics (QED).
First, we give the description of the cross-Kerr effect induced
between two resonators in circuit QED. Then we use the cross-Kerr
media to design the effective quantum nondemolition (QND)
measurement on microwave-photon number. By using the QND
measurement, the parties in quantum communication can accomplish the
entanglement concentration and purification of nonlocal two-mode
squeezed microwave photons. We discuss the feasibility of our
schemes by giving the detailed parameters which can be realized with
current experimental technology. Our work can improve some practical
applications in continuous-variable microwave-based quantum
information processing.
\end{abstract}


\maketitle

\section{Introduction}

Nowadays, quantum entanglement plays an indispensable role in
quantum communication, such as quantum teleportation
\cite{CHBennet1993}, quantum dense coding
\cite{CHBennet1992,superdense}, quantum key distribution
\cite{AKEkert,bbm92}, quantum secret sharing \cite{MHillery}, and
quantum secure direct communication
\cite{longliupra,FGDeng2003,twostepexp,twostepexp2}. To realize
quantum communication, the legitimate parties should first construct
a quantum channel. This critical  step usually requires nonlocally
maximally entangled states between two  remote parties to promise an
high-efficiency quantum communication. However, it is hard for the
parties to keep a nonlocally maximally entangled state due to the
decoherence from the environment in the process of the transmission
and storage of the states in practice. To overcome this problem,
some effective approaches have been proposed, such as the
error-rejecting coding with decoherence free subspaces
\cite{DFS1,DFS2,DFS3}, entanglement concentration
\cite{CHBennett1996,YBSheng2008,BCRen012302,Lixhpraecp,Lixhpraecp2,caocecpoe,caocecpaop,Renbcaop,HZhangPRA2017},
 and entanglement purification
\cite{EPP1,DDeutschPRL1996,JWPannature2001,Simon2002,YBShengPRA2008,EPP2,EPP3,EPP4,DEPPDeng,EPP5,EPPwanggy,DuFFHEPP,EPPwangc,EPPadd1,EPPadd2,EPPadd3,EPPadd4,zhanghaoepp}.

As another important form of quantum entanglement, continuous
variable systems have been used for quantum information processing
(QIP) \cite{BraunsteinRMP77513}, such as continuous variable
teleportation \cite{VaidmanPRA1473,BraunsteinPRL869}, continuous
variable quantum computation \cite{LloydPRL1784}, error correction
\cite{BraunsteinNATURE394,LloydPRL4088}, and  continuous variable
quantum cryptography \cite{RalphPRA010302}. Continuous-variable
quantum systems have the advantages of cheap resources and easy
generation and control in QIP. Similar to the discrete variable,
continuous variable systems also suffer from the decoherence
inevitably. Therefore, some interesting methods are proposed to
overcome this problem, such as continuous-variable entanglement
concentration
\cite{LMDuanPRL2000,LMDuanPRA2000,Fiurasekpra022304,Menziespra042315,Menziespra062310,Tathampra14},
purification and distillation
\cite{LMDuanPRL2000,LMDuanPRA2000,Opatrnypra032302,Olivarespra032314,Dattaprl060502,Eisertny431,Kitagawapra042310,lundpra032309,Ourjoumtsevprl030502,HageNP915,DongNP919,
Takahashinp178}. For example, in 2000, Duan \emph{et al.}
\cite{LMDuanPRL2000} proposed an efficient entanglement
concentration and purification protocol for continuous-variable
quantum systems. In 2012, Datta \emph{et al.} \cite{Dattaprl060502}
proposed the scheme for a compact continuous-variable entanglement
distillation. In 2007, Ourjoumtsev \emph{et al.}
\cite{Ourjoumtsevprl030502} experimentally increased the
entanglement between Gaussian entangled states by non-Gaussian
operations. In 2008, Hage \emph{et al.} \cite{HageNP915} prepared
the distilled and purified continuous-variable entangled states in
experiment. In the same year, Dong \emph{et al.} \cite{DongNP919}
experimentally realized the entanglement distillation of mesoscopic
quantum states. In 2010, Takahashi \emph{et al.}
\cite{Takahashinp178} realized the entanglement distillation from
Gaussian input states.


Circuit quantum electrodynamics (QED), which couples the
superconducting qubit to transmission line resonators \cite{ABlais},
is a very good platform for operating the interaction between a
superconducting qubit and a microwave photon. With the advantages of
good tunability and scalability, circuit QED has already been
studied widely for QIP
\cite{ABlais2,DiCarlo,LongcircuitPRA,Wangsuperconducting,circuitTianlPRL,3q,3q1,MHofheinz,
DISchuster,BRJohnson,CPYangPRA2013,HuaMPRA,SongXK1,SongXK2,zhaopra053,Kellynature66,CSong10bit}.
The realization of 9-qubit state preservation \cite{Kellynature66}
and 10-qubit \cite{CSong10bit} entanglement in experiment indicates
that the superconducting qubit has a big potential in quantum
computation. On the other hand, the manipulation of a microwave
quantum state is also a meaningful research area in circuit QED.
Cross-Kerr effect, a typical nonlinear effect, has been studied both
theoretically \cite{SRebic2009,YHu,TLiuQIP} and experimentally
\cite{ICHoi,ETHolland} in recent years. In 2011, Hu \emph{et al.}
\cite{YHu} proposed the cross-Kerr effect induced by coupling
resonators to a superconducting molecule in circuit QED. In 2015,
Holland \emph{et al.} \cite{ETHolland} demonstrated this cross-Kerr
effect between two resonators in experiment. Due to the strong
anti-interference and low loss in the process of transmission,
microwave photon becomes a very important flying bit in both
classical and quantum communication. However, microwave photons also
cannot avoid the decoherence in quantum communication and will
change from a maximally quantum entangled state to a partially
entangled pure state or a mixed one. Therefore, the entanglement
concentration and purification of microwave quantum state are
indispensable for promising an effective microwave-based quantum
communication. For continuous-variable microwave quantum systems,
there is no research in this area.

In this paper, we propose the first physical implementation scheme
for the entanglement concentration and purification of two-mode
squeezed microwaves, one kind of continuous-variable systems, in
circuit QED. Using our scheme, the parties in quantum communication
can effectively concentrate and purify the pure and mixed two-mode
squeezed microwaves in long distance microwave quantum
communication, respectively. Superconducting
circuit is easy to operate the microwave-based QIP with current
experimental technology due to its good tunability. Our scheme will
improve the applications in nonlocal microwave-based quantum
communication with continuous-variable quantum states.

This paper is organized as follows: In Sec.~\ref{sec2}, we introduce
the physical implementation of cross-Kerr effect and QND measurement
of photon number of microwave photon in two cascade resonators in
circuit QED.
In Sec.~\ref{sec3} and Sec.~\ref{sec4}, we perform the physical
implementation for the entanglement concentration and purification
of two-mode squeezed state of microwave, respectively. A discussion
and a summary are given in Sec.~\ref{sec5}.

\section{Quantum nondemolition measurement based on Kerr effect in circuit QED}  \label{sec2}

\begin{figure}[!ht]
\centering\includegraphics[width=8cm]{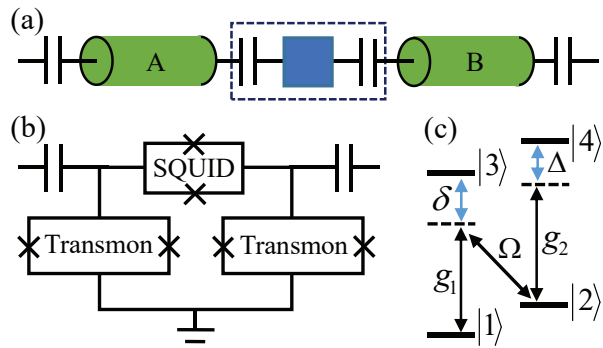}
\caption{(Color online) Schematic diagram for the cross-Kerr effect in circuit QED.
(a) Cross-Kerr effect induced by coupling two resonators to a
superconducting molecule. (b) The detailed structure of
superconducting circuit in the dashed box above. The symbols of
cross represent the Josephson junctions. (c) The level structure of
the molecule and the corresponding interactions \cite{YHu}.}
\label{crosskerr}
\end{figure}

The cross-Kerr effect can be induced by coupling two resonators to a
four-level superconducting molecule in circuit QED. The schematic
diagram is shown in Fig.~\ref{crosskerr}(a). A and B are two
resonators, and the middle box is a four-level superconducting
molecule constructed by two transmon qubits \cite{JKoch} shown in
Fig.~\ref{crosskerr}(b). The level structure and corresponding
couplings are described in Fig.~\ref{crosskerr}(c). The resonators A
and B couple to $|1\rangle-|3\rangle$ and $|2\rangle-|4\rangle$,
respectively. The corresponding coupling strengths are $g_{1}$ and
$g_{2}$, respectively. The detunings are $\delta$ and $\Delta$,
respectively. The classical field with the strength $\Omega$ is
resonant with level $|2\rangle-|3\rangle$. In the interaction
picture, the Hamiltonian is given by \cite{YHu} (with $\hbar=1$)
\begin{eqnarray}        \label{eq2}
\hat{H}\!=\!\delta\hat{\sigma}_{33}+\Delta\hat{\sigma}_{44}
+i\text{g}_{1}\left(\hat{\sigma}_{13}\hat{a}^{\dag}-\hat{\sigma}_{31}\hat{a}\right)+i\text{g}_{2}\left(\hat{\sigma}_{24}\hat{b}^{\dag}-\hat{\sigma}_{42}\hat{b}\right)
+i\Omega\left(\hat{\sigma}_{23}-\hat{\sigma}_{32}\right).
\end{eqnarray}
Here the detunings $\delta=E_{31}-\omega_{a}$ and
$\Delta=E_{42}-\omega_{b}$. The level spacing $E_{ij}$ is defined
with $E_{ij}=E_{i}-E_{j}$. $\omega_{a}$ and $\omega_{b}$ are the
frequencies of resonators A and B, respectively. $\hat{a}$
$(\hat{a}^{\dag})$ and $\hat{b}$ $(\hat{b}^{\dag})$ are the
annihilation (creation) operators for resonators A and B,
respectively. The operator $\hat{\sigma}_{ij}$ is defined with
$\hat{\sigma}_{ij}=|i\rangle\langle j|$. When the parameters satisfy
the conditions with  $|\text{g}_{1}/\Omega_{c}|^{2}\ll 1$ and
$|\text{g}_{2}|\ll|\Delta|$ \cite{AImamoglu1997}, one can obtain the
effective cross-Kerr Hamiltonian \cite{YHu}
\begin{eqnarray}        \label{eq3}
\hat{H}_{Kerr}=\chi\hat{a}^{\dag}\hat{a}\hat{b}^{\dag}\hat{b},
\end{eqnarray}
where
$\chi=-\text{g}_{1}^{2}\text{g}_{2}^{2}/(\Delta\Omega^{2}_{c})$ is
the cross-Kerr coefficient.

The four-level artificial molecule \cite{YHu} used before (dashed
box in Fig.~\ref{crosskerr}(a)) can be constructed with two transmon
qubits \cite{JKoch} and a superconducting quantum interference
device (SQUID) \cite{JSiewert}. The detailed structure is shown in
Fig.~\ref{crosskerr}(b). The top loop represents SQUID and the two
bottom loops are transmon qubits. The crosses in each loop are
Josephson junctions. The two transmon qubits can couple to each
other via the SQUID. By using the two level language, the coupling
system is translated to a superconducting molecule \cite{YHu,JKoch}
with four levels shown in Fig.~\ref{crosskerr}(c).

\begin{figure}[!ht]
\includegraphics[width=10cm]{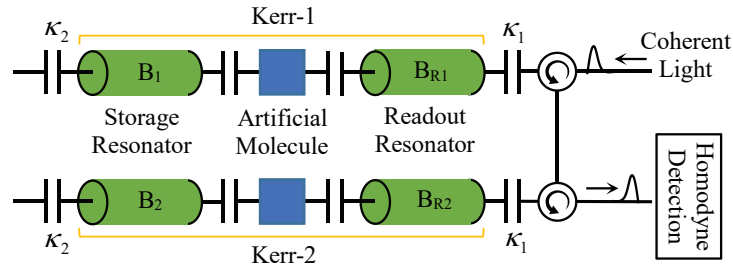}
\centering\caption{(Color online) Schematic diagram for the QND measurement by using
cross-Kerr effect in circuit QED. The two cross-Kerr media are the
same ones. $B_{1}$ and $B_{2}$ are the storage resonators. $B_{R1}$
and $B_{R2}$ are the readout resonators} \label{QND}
\end{figure}

In the schemes of entanglement concentration and purification, the QND
measurement system is usually a crucial part. Here, we use the two
same cross-Kerr media induced by circuit QED to realize the QND
measurement shown in Fig.~\ref{QND}. We choose the storage and
readout resonators with low and high decay rates, respectively. When
the probe light is resonant with readout resonators, the
Heisenberg-Langevin equations for two same cross-Kerr media are
given by
\begin{eqnarray}\label{HLequation}
\frac{d\hat{a}_{i}}{dt}=-i\chi\,\hat{n}_{i}\hat{a}_{i}-\frac{\kappa_{1}}{2}\hat{a}_{i}-\sqrt{\kappa_{1}}\,\hat{a}^{in}_{i},(i=1, 2).
\end{eqnarray}
Here, the input of the first resonator is
$\hat{a}^{in}_{1}=g\sqrt{\kappa_{1}}$ (without noise), where
$g\sqrt{\kappa_{1}}$ is a constant driving field. As a cascade
system, the input of the second resonator equals the output of
the first resonator with the formula
$\hat{a}^{in}_{2}=\hat{a}^{out}_{1}$. We assume the decay rates of
readout resonators are very large and satisfy $\kappa_{1}\gg\chi$,
after adiabatically eliminating the cavity modes
$(\dot{\hat{a}}_{i}=0)$, we can get the output field of the second
resonator as
\begin{eqnarray}        \label{ouput2}
\hat{a}^{out}_{2}\approx-\frac{4ig\chi}{\sqrt{\kappa_{1}}}(\hat{n}_{1}+\hat{n}_{2})+g\sqrt{\kappa_{1}}.
\end{eqnarray}
Here we use the standard input-output equation
$\hat{a}_{out}=\hat{a}_{in}+\sqrt{\kappa_{1}}\,\hat{a}$ in
calculation.

When we make a homodyne measurement on the $X$ component of the
quadrature phase amplitudes of the output field of the second
readout resonator $\hat{a}^{out}_{2}$, the measuring operator is
$\hat{X}(\tau)=\frac{1}{\tau}\int^{\tau}_{0}\frac{1}{\sqrt{2}}[\hat{a}^{out}_{2}(t)+\hat{a}^{out\dag}_{2}(t)]dt$.
Here, $\tau$ is the measuring time. Substituting the result of
$\hat{a}^{out}_{2}$ into the measuring operator and choosing
$g=i|g|$, we can get the result as
\begin{eqnarray}        \label{measurement}
\hat{X}(\tau)=\frac{4\sqrt{2}|g|\chi}{\sqrt{\kappa_{1}}}(n_{1}+n_{2}).
\end{eqnarray}
The signal is proportional to the total photon number $n_{1}+n_{2}$.
One can infer the total photon number according to the signal.

\section{The entanglement concentration of two-mode squeezed microwave states} \label{sec3}

\begin{figure}[!ht]
\centering\includegraphics[width=14.0cm]{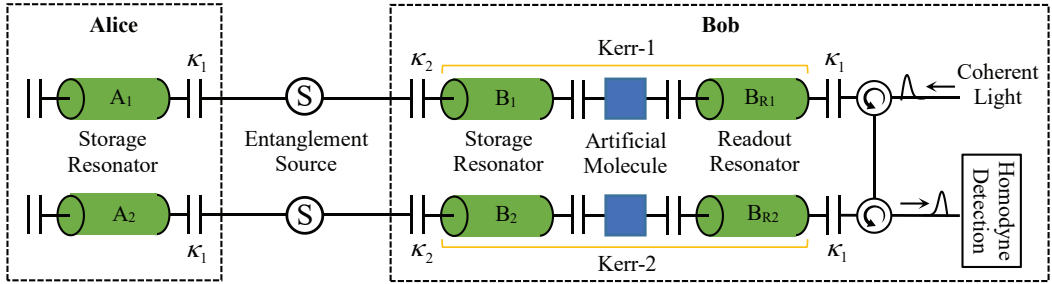}
\caption{(Color online) Schematic diagram for the entanglement concentration of
two-mode squeezed microwave photons. The two remote parties, say
Alice and Bob, hold the resonators $A_{1}A_{2}$ and $B_{1}B_{2}$,
respectively. The same two-mode squeezed states are prepared between
superconducting resonators on two sides. Bob holds the QND
measurement system composed of two same cross-Kerr media. The probe
light input from resonator $B_{R1}$ and will be detected via a
homodyne detection after it leaves $B_{R2}$. The circle with an arrow
is a circulator.} \label{concentration}
\end{figure}

Here, we perform the physical implementation for entanglement
concentration protocol of two-mode squeezed microwaves. We choose
the theoretic entanglement concentration protocol proposed by Duan
\emph{et al.} \cite{LMDuanPRL2000}. The detailed schematic diagram
is shown in Fig. \ref{concentration}. Here, we consider Alice and
Bob hold the resonators $A_{1}A_{2}$ and $B_{1}B_{2}$, respectively.
$A_{i}B_{i}$ $(i=1,2)$ are prepared in two-mode squeezed microwave
states
$|\psi\rangle_{AB}=exp[r(\hat{a}^{\dag}_{A}\hat{a}^{\dag}_{B}-\hat{a}_{A}\hat{a}_{B})]|0\rangle$
with the form expanded in Fock state basis
\begin{eqnarray}    \label{squeezed}
|\psi\rangle_{AB}=\sqrt{1-\lambda^{2}}\sum^{\infty}_{n=0}\lambda^{n}|n,n\rangle_{AB},
\end{eqnarray}
where $\lambda=tanh(r)$  and $r$ is the squeezing parameter. The
magnitude of entanglement of this pure squeezed state is given by
\begin{eqnarray}    \label{entanglesqueezed}
E(|\psi\rangle_{AB})\!=\!cosh^{2}(r)ln[cosh^{2}(r)]\!-\!sinh^{2}(r)ln[sinh^{2}(r)].
\end{eqnarray}
So, the state of this composite system composed of two pairs of
squeezed states can be written as \cite{LMDuanPRA2000}
\begin{eqnarray}    \label{squeezed}
|\psi\rangle_{1}&=&|\psi\rangle_{\!A_{1}\!B_{1}}\otimes|\psi\rangle_{\!A_{2}\!B_{2}}\nonumber\\
&=&(1-\lambda^{2})\sum^{\infty}_{m=0}\lambda^{m}\sqrt{1+m}|m\rangle_{\!A_{1}\!A_{2}\!B_{1}\!B_{2}},
\end{eqnarray}
where the state $|m\rangle_{\!A_{1}\!A_{2}\!B_{1}\!B_{2}}$ is
\begin{eqnarray}    \label{}
|m\rangle_{\!A_{1}\!A_{2}\!B_{1}\!B_{2}}\!=\!\frac{1}{\sqrt{1\!+\!m}}\!\sum^{m}_{n=0}\!|n,m\!-\!n\rangle_{\!A_{1}\!A_{2}}|n,m\!-\!n\rangle_{\!B_{1}\!B_{2}}.
\end{eqnarray}

Now, Bob makes a local QND measurement on the total photon number of
the cavities $B_{1}$ and $B_{2}$. When Bob gets the result with $m$,
the state $|\psi\rangle_{1}$ will collapse to
$|m\rangle_{\!A_{1}\!A_{2}\!B_{1}\!B_{2}}$ with the  probability
$p_{m}=[(1-\lambda^{2})\lambda^{m}]^{2}(1+m)$. The magnitude of
entanglement of $|m\rangle_{\!A_{1}\!A_{2}\!B_{1}\!B_{2}}$ is
\begin{eqnarray}    \label{}
E(|m\rangle_{\!A_{1}\!A_{2}\!B_{1}\!B_{2}})=ln(1+m).
\end{eqnarray}
If
$E(|m\rangle_{\!A_{1}\!A_{2}\!B_{1}\!B_{2}})>E(|\psi\rangle_{AB})$,
Alice and Bob get the state with more entanglement. According this
inequality, one can see that $m$ should satisfy the requirement
\begin{eqnarray}    \label{}
m>\frac{[cosh^{2}(r)]^{cosh^{2}(r)}}{[sinh^{2}(r)]^{sinh^{2}(r)}}-1.
\end{eqnarray}
Therefore, if the result of the QND measurement satisfies the above
inequality, Alice and Bob can keep the corresponding maximally
 entangled microwave state.

\section{The entanglement purification of two-mode squeezed microwave states} \label{sec4}

\begin{figure}[!ht]
\centering\includegraphics[width=17.0cm]{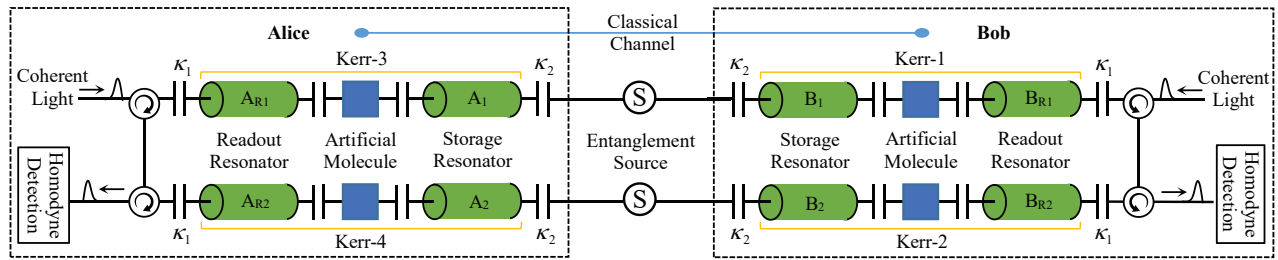}
\caption{(Color online) Schematic diagram for the entanglement purification of
two-mode squeezed microwave photons.  Alice and Bob hold the
$A_{1}A_{2}$ and $B_{1}B_{2}$, respectively. Both Alice and Bob have
QND measurement systems. The Kerr-1 (Kerr-3) and Kerr-2 (Kerr-4) are
the same ones. The classical communication channel is used to
compare the measuring results from Alice and Bob. The circle with an
arrow is a circulator.} \label{purification}
\end{figure}

In practice, one cannot avoid the noise in the process of the state
preparation. It may appear the mixed state due to the influence of
noise. Therefore, we need purify the mixed entangled state for
high-fidelity quantum communication. The detailed schematic diagram
of entanglement purification is shown in Fig. \ref{purification}.
Compared with the concentration process, the difference here is that
both Alice and Bob hold the QND measurement systems. The classical
channel is used to compare the measurement results. Now, we consider
the situation with a very small noise. According to the quantum
trajectory theory, the state of two pairs can approximatively be
expressed in two situations. If there are no jumps, the state is
\cite{LMDuanPRL2000,LMDuanPRA2000}
\begin{eqnarray}    \label{}
|\psi\rangle_{no}\!=\!\frac{1\!-\!\lambda^{2}}{\sqrt{p_{no}}}\!\sum^{\infty}_{m=0}\!\lambda^{m}e^{-\eta\tau m/2}\sqrt{1+m}|m\rangle_{\!A_{1}\!B_{1}\!A_{2}\!B_{2}}.
\end{eqnarray}
Here, the probability is
$p_{no}=\frac{(1-\lambda^{2})^{2}}{(1-\lambda^{2}e^{-\eta\tau})^{2}}$.
The total loss rate $\eta$ is $\eta=\eta_{A}+\eta_{B}$ and $\tau$
is the transmission time. When the jump occurs, the state becomes
\cite{LMDuanPRL2000,LMDuanPRA2000}
\begin{eqnarray}    \label{mixed}
|\psi\rangle^{x_{i}}_{jump}\!=\!\sqrt{\frac{\eta_{x}\tau}{p^{x_{i}}_{jump}}}a_{x_{i}}|\psi\rangle_{\!A_{1}\!B_{1}}\otimes|\psi\rangle_{\!A_{2}\!B_{2}},
\end{eqnarray}
where $x=A,B$ and $i=1,2$. The probability is
$p^{x_{i}}_{jump}=\bar{n}\eta_{x}\tau$ with the mean photon number
of single mode $\bar{n}=sinh^{2}(r)$. Here, we consider that the
quantum jump only occurs on one side of Alice and Bob. Both Alice
and Bob should make a QND measurement on their two resonators. When
they get the same result with $m_{A}=m_{B}=m$ compared by classical
channel, they keep this maximally entangled state with entanglement
$ln(1+m)$. If the result of photon number is different, they should
discard this situation.

\section{discussion and summary}\label{sec5}

The cross-Kerr effect in circuit QED should be realized with
reasonable parameters. According to the previous work \cite{YHu}, we
can choose the parameters as follows. To satisfy the conditions
$|\text{g}_{1}/\Omega_{c}|^{2}\ll 1$, $|\text{g}_{2}|\ll|\Delta|$,
the coupling strengths are chosen as
$\text{g}_{1}/2\pi\sim\text{g}_{2}/2\pi\sim 300$ MHz. The detuning
and the strength of classical pump field are chosen with
$\Delta/2\pi\sim\Omega_{c}/2\pi\sim 1.5$ GHz. With the above
parameters, the strength of cross-Kerr effect in our scheme is
$|\chi|/2\pi\sim 2.4$ MHz. In the recent circuit QED experiment \cite{ETHolland}, the first single-photon-resolved cavity-cavity cross-Kerr interaction has been observed with a state dependent shift $ \chi_{sc}/2\pi=2.59\pm0.06$ MHz.
For two-mode squeezed states, we choose
the squeezed parameter with $r\sim0.9$ and the mean photon number
$\langle n\rangle=sinh(r)\sim1.1$. We choose $|g|\sim50$. To keep an
effective QND measurement, the condition $\kappa_{1}\gg\chi$ should
be met. The decay rates of readout and storage resonator are set
with $\kappa_{1}/2\pi\sim 100$ MHz and $\kappa_{2}/2\pi\sim 20$ kHz,
respectively.

In practice, the QND system will be influenced with the noise from
environment. When we consider the standard vacuum white noise, we
rewrite the input field with
$\hat{a}^{in}_{1}\!=\!g\sqrt{\kappa_{1}}+\hat{a}^{in'}_{1}$, where
the noise term $\hat{a}^{in'}_{1}$ satisfies the relations
$\langle\hat{a}^{in'\dag}_{1}(t)\hat{a}^{in'}_{1}(t')\rangle=0$ and
$\langle\hat{a}^{in'}_{1}(t)\hat{a}^{in'\dag}_{1}(t')\rangle=\delta(t-t')$.
Then the noise term will make a contribution to the measurement
result. According to previous work \cite{LMDuanPRA2000}, promising
an effective QND measurement should satisfy the requirement
$\kappa_{1}/(64|g|^{2}\chi^{2})<\tau<1/\kappa_{2}$. With the
parameters given above, the measuring time should be $0.02$ ns
$<\tau<8$ $\mu$s in our system.
The imperfections in the QND measurement also influence the protocol. Here, to make an effective measurement, we show the requirements for some parameters given in the previous work \cite{LMDuanPRA2000}. When the phase of driving field is unstable, i.e., $g$ is $g=i|g|e^{i\delta}$. The phase instability should satisfy $\delta<4\chi/\kappa_{1}$. If the decay rates and the Kerr coefficients for the resonators are not identical, we denote the decay rates for readout resonators $B_{R1}$ and $B_{R2}$ with $\kappa_{R1}$ and $\kappa_{R2}$, respectively. The Kerr coefficients for Kerr-1 and Kerr-2 are labeled with $\chi_{1}$ and $\chi_{2}$. At this point, one should keep them with $|\frac{\chi_{2}\kappa_{R1}}{\chi_{1}\kappa_{R2}}-1|<1/\langle n_{B2}\rangle$. Actually, there exist the absorption and leakage of the probe light in some devices, such as circulators and resonators. We use $\gamma_{i}(i=1,2)$ to represent all the loss and the $\gamma_{i}$ should follow the requirement $\gamma_{i}<\kappa_{1}/\langle n_{Bi}\rangle^{2}$.

In summary, we have presented the first physical implementation of
entanglement concentration and purification protocol for two-mode
squeezed microwave photons in circuit QED. The protocol can be extended to
multiple entangled pairs with adding more cascade QND measurement
systems. Our scheme has the advantage that it can be realized easily in practice with current
experimental technology. Our work will improve the feasibility of
nonlocal microwave-based quantum communication with
continuous-variable quantum states.

\section*{ACKNOWLEDGMENTS}
This work is supported by the National Natural Science Foundation of
China under Grants Nos. 11474026 and 11674033, and the
Fundamental Research Funds for the Central Universities under Grant
No. 2015KJJCA01.

\end{document}